# Unified interface dipole theory for Fermi level pinning effect at metal-semiconductor contacts

Ziying Xiang[1,2], Jun-Wei Luo[1,2,*], and Shu-Shen Li[1,2]

[1]*State Key Laboratory of Semiconductor Physics and Chip Technologies, Institute of Semiconductors, Chinese Academy of Sciences, Beijing 100083, China*

[2]*Center of Materials Science and Optoelectronics Engineering, University of Chinese Academy of Sciences, Beijing 100049, China*

*Email: jwluo@semi.ac.cn.

## ABSTRACT

We present a unified bond dipole theory for metal-semiconductor interfaces to explain the microscopic origin of interface dipoles and Fermi level pinning (FLP) in terms of Harrison's bond-orbital model. By combining first-principles calculations with tight-binding analysis, we show that localized bonding between semiconductor surface dangling bonds and metal orbitals is sufficient to generate a large interface dipole and induce strong FLP, even when only a single metal monolayer is present. Within this framework, metal-induced gap states (MIGS), dangling-bond-induced surface states (DBSS), and bonding states embedded in the valence band are all understood as different outcomes of the same underlying interface bonding mechanism, rather than as independent causes of FLP. We further establish that the key parameter governing FLP strength is the density of surface dangling bonds that can form new chemical bonds with the metal, which directly controls the magnitude of the bond-induced interface dipole. This picture naturally explains the weaker pinning observed in more ionic semiconductors than in covalent ones and provides practical guidance for engineering metal-semiconductor interfaces and tuning Schottky barrier heights.

# I. INTRODUCTION

The Fermi level pinning effect induces a large Schottky barrier height (SBH) at metal-semiconductor (MS) interfaces in semiconductors, especially in Si and Ge, posing a major challenge to reducing contact resistance and achieving high-performance devices [1-3]. For n-type semiconductors, the SBH (N-SBH) $\Phi_{B,n}$ is expected to be given by the difference between the metal work function (WF) $\phi_M$ and the semiconductor electron affinity (EA) $\chi_S$ according to the Schottky-Mott rule [4,5]:

$$\Phi_{B,n} = \phi_M - \chi_S \tag{1}$$

implying that the SBH can be reduced by choosing metals with a sufficiently lower $\phi_M$. However, experimental observations show that N-SBHs are insensitive to changes in the contacting metals and thus difficult to modulate, as the Fermi levels are pinned nearly at a specific energy level, which is often referred to as the charge neutrality level (CNL) [6], within the semiconductor band gap. This notorious phenomenon is known as Fermi level pinning (FLP) [7], with its strength characterized by the pinning factor $S$:

$$S = \frac{d\Phi_{B,n}}{d\phi_M}. \tag{2}$$

For $S = 0$, the Fermi level is fully pinned and corresponds to the Bardeen limit [8], while $S = 1$ indicates no pinning and corresponds to the Schottky-Mott limit [4,5]. FLP is frequently regarded as being caused by the interface dipole potential $\Delta V$, which originates from the charge transfer and rearrangement between the metal and semiconductor, modifying the N-SBH $\Phi_{B,n}$ as follows [7]:

$$\Phi_{B,n} = \phi_M - \chi_S - \Delta V. \tag{3}$$

Differentiating both sides of Eq. (3) with respect to $\phi_M$, and assuming $\chi_S$ to be a constant, one can rewrite Eq. (2) as:

$$S = \frac{d\Phi_{B,n}}{d\phi_M} = 1 - \frac{d\Delta V}{d\phi_M}. \tag{4}$$

One can see that if $\Delta V$ fully compensates for the change in $\phi_M$, such that $k \equiv \frac{d\Delta V}{d\phi_M} \to 1$, the resulting interface dipole completely offsets metal work function variations, yielding the extremely strong FLP observed in materials like Ge and Si [8,9].

Despite considerable progress made in the past century, the origin of the charge transfer across the interface that generates the interface dipole remains under debate [10-15]. Currently, the most widely accepted model is the gap states model [15-18]. Cowley and Louie [10,15,18] quantitatively attributed the interface dipole to the occupation of gap states in a density $D_s$ (number of states per area per eV) with energy levels below the Fermi level but above the CNL $\phi_{\text{CNL}}$. The induced interface charge $eD_s(\Phi_{B,n} - \phi_{\text{CNL}})$ is balanced by the opposite charge with the same magnitude on the metal side, as required by the charge neutrality, creating an interface dipole and thus inducing an interface potential drop $\Delta V$,

$$\Delta V = -(\Phi_{B,n} - \phi_{\text{CNL}}) \frac{e^2 D_s \delta_{\text{it}}}{\varepsilon_{\text{it}}}. \tag{5}$$

Here, $\varepsilon_{\text{it}}$ is the dielectric constant in the interface region, and $\delta_{\text{it}}$ is the effective distance of the interface gap states determined by their decay length, i.e., the distance over which the wave function amplitude decreases exponentially into a material. We assume $\Delta V$ to be positive when electrons transfer from the semiconductor to the metal. Therefore, according to the gap states model, the pinning factor $S$ is related to the gap states as follows [18]:

$$S = \left(1 + \frac{e^2 D_s \delta_{\text{it}}}{\varepsilon_{\text{it}}}\right)^{-1}. \tag{6}$$

In previous reports, various types of gap states associated with the FLP have been proposed, including dangling-bond (DB) induced surface states [13], metal-induced gap states (MIGS) [1], and interface defect states [11]. MIGS are roughly free-electron-like metal wave functions penetrating into the semiconducting side, where the penetration depth is inversely proportional to the size of the semiconductor band gap [19]. Because dangling bonds are believed to be passivated upon contact formation, it has been commonly argued that MIGS dominate the interface states responsible for the interface dipole for FLP [1].

Although the gap states model has been widely used and appears to account for the different pinning strengths observed at various MS interfaces [15,20], recent experiment have raised doubts about its completeness [21]. It demonstrated that charge from interface bonding states outside the semiconductor band gap and deep within the valence bands could also contribute to the interface dipole [21], in conflict with the gap states model, which accounts exclusively for those interface states within the semiconductor band gap [15,20]. Furthermore, Tung recently demonstrated in experimental measurements that the orientations and terminations of the semiconductor surface can also alter SBH [22], in contrast to the gap states model given in Eq. (6), according to which FLP is entirely governed by the intrinsic property of bulk semiconductors [6,9]. In response to these limitations, an alternative perspective has emerged by emphasizing the role of interface bonding in the formation of interface dipoles. Early studies by Harrison, Tersoff, and Lambrecht highlighted the significance of chemical bonding in band alignment [14,23-25], followed later by Tung who proposed a bond polarization model in an attempt to explicitly link the bond polarization to the FLP effect [12,26]. These theoretical models have been supported by both experimental and computational results [21,27], providing a complementary perspective on the origin of interface dipoles. Nevertheless, these models have not yet established a robust connection between interface chemical bonding, gap states, and the strong FLP effect. The validity of the gap states model therefore remains an open question, and a unified framework to explain the formation and behavior of interface dipoles is still lacking.

In this work, we identify that MIGS originates from the localized interface bonding between semiconductor surface dangling bonds and metal atoms. We demonstrate that both MIGS and dangling-bond-induced surface states (DBSS) can be rooted in the same physical origin, allowing them to be unified under the interface bonding framework. We also develop a bond dipole theory for MS interfaces based on the bond-orbital model proposed by Harrison [28], which illustrates that chemical bond polarization at the interface alone generates a dipole potential that accounts for the strong FLP observed in Si and Ge and explains FLP from the perspective of interface bonding.

## II. RESULTS AND DISCUSSION

### A. Identify the origin of MIGS as the interface bonding states

It has been well established that Ge exhibits the strongest FLP effect among all semiconductors in experiments, with a pinning factor $S \approx 0.02$ approaching the Bardeen limit [9,29], but its origin remains under debate. In the context of conventional theories, the FLP effect is caused by charge rearrangement at the MS interface due exclusively to metal-induced gap states (MIGS) [1,10,15,18], which originate from the penetration of metal Bloch wave functions into the semiconductor band gap. However, our first-principles calculations have revealed another type of interface gap states, which are localized within 1~2 atomic layers at the MS interface and decay exponentially on both sides [30], in sharp contrast to MIGS that are bulk-like inside the metal and rapidly decay into the semiconductor. These strongly localized states at the MS interface emerge from the dangling bonds by the surface truncation and are thus referred to as dangling-bond-induced surface states (DBSS). We have also illustrated that Ge prefers a non-reconstruction interface, whereas Si prefers a reconstructed interface at MS contacts [30]. The reconstruction of interface Si atoms yields a self-passivation of interface dangling bonds to reduce the density of DBSS [30], which is a predominant factor over the MIGS responsible for a weaker FLP at metal-Si contacts ($S \approx 0.16$ [32,33]) than at metal-Ge contacts ($S \approx 0.02$ [9,29]). To address the origin of MIGS, in this work we first examine the effect of the thickness of the metal layer on FLP at MS contacts. Figure 1 shows the first-principles calculated N-SBHs of metal-Ge (001) and metal-Si (001) contacts in the thick-layer limit (with pinning factor $S = 0$ and $S = 0.16$, respectively), in comparison with the cases of a single monolayer metal adsorbed on the Ge (001) and Si (001) surfaces. The monolayer contacts exhibit pinning factors of $S = 0.08$ for Ge and $S = 0.19$ for Si, very close to those of the corresponding thick-layer contacts, indicating that even a single monolayer of metal is sufficient to induce strong FLP. This, in turn, implies that both thick-layer and monolayer metal models share the same underlying mechanism responsible for FLP at MS contacts.

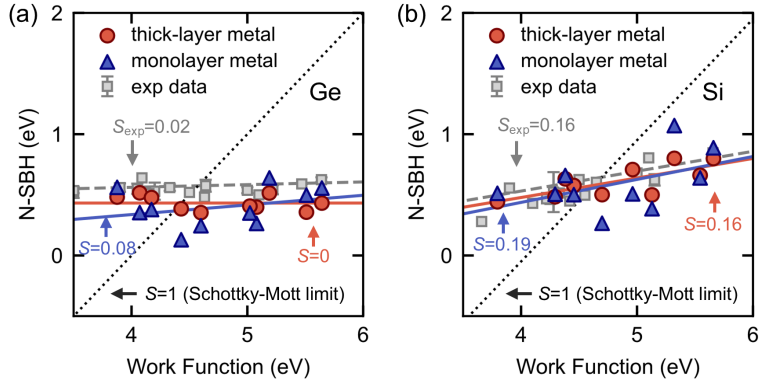

**FIG. 1.** First-principles calculated n-type Schottky barrier heights (N-SBHs) for (a) metal-Ge (001) and (b) metal-Si (001) contacts. In each panel, both thick-layer metal contacts [30] and monolayer metals adsorption on the semiconductor surface are compared with experimental N-SBHs for conventional thick-layer metal-semiconductor contacts [9,29,32-37] and with the ideal Schottky-Mott limit ($S = 1$).

Figure 2 shows the first-principles band structures for three sequential stages of Ag-Ge contact formation: (i) a bare Ge (001) surface, (ii) a single Ag monolayer deposited on the Ge (001) surface, and (iii) a thick-layer of Ag deposited on the Ge (001) surface with an ideal non-reconstructed atomic structure. Figure 2(a) shows that the bare Ge (001) surface exhibits periodic Ge dangling bonds (DBs), which generate two surface bands with surface states highly localized at the surface and energy levels distributed across the whole band gap. The bands with small energy dispersion (marked in blue) derive from the atomic $p_z$ orbitals of the surface DBs, and the bands with large energy dispersion (red) derive from the hybridization of the $p_x$ and $p_y$ orbitals of the surface DBs. Upon adsorption of a single monolayer of Ag, the $p_z$-derived DB bands are strongly disrupted because of bonding with metal states, whereas the $p_x/p_y$-derived DB bands are barely perturbed, maintaining their original DB character, as simultaneously highlighted by the blue and cyan dots in Fig. 2(b). The subsequent deposition of additional Ag layers further modulates the disrupted $p_z$-derived DB bands by splitting them into multiple sub-bands, as shown in Fig. 2(c), and reduces the density of states inside the band gap (as quantified by the smaller dots), whereas the $p_x/p_y$-derived DB bands remain nearly unchanged.

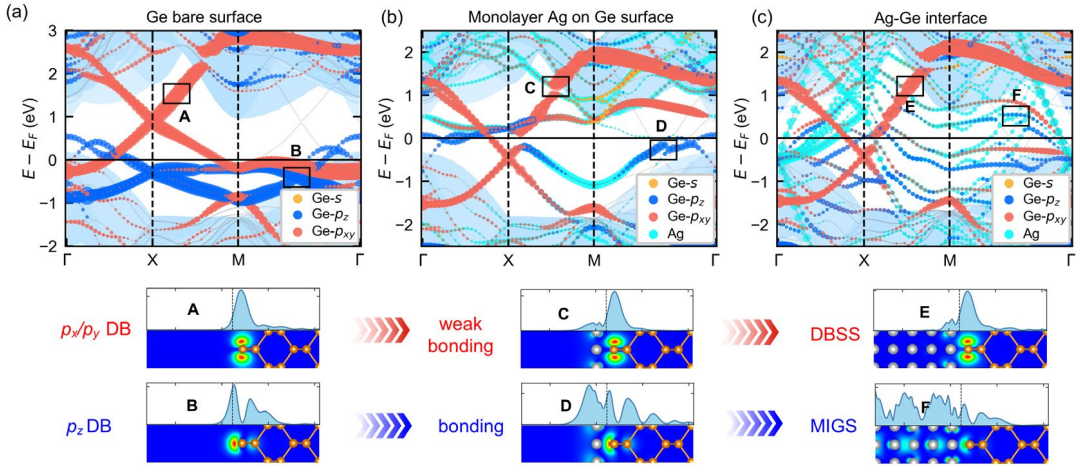

**FIG. 2.** First-principles band structures for three sequential stages: (a) clean Ge surface, (b) Ge surface covered by a monolayer of Ag atoms, and (c) Ag-Ge contact formation. In the top panels, the light-blue shadow represents the bulk Ge band structure projected onto the (001) surface Brillouin zone, the gray lines denote the interface band structure and colored dots with varying sizes represent the contributions from Ge and Ag. The bottom panels show the partial charge distributions for representative interface states. Orange spheres denote Ge atoms, and silver spheres denote Ag atoms. The black dashed lines mark the center between the two surfaces.

The bottom panel of Fig. 2(c) shows the partial charge densities of these two sets of surface DB bands. It reveals that the DB states of the perturbed bands have a continuously propagating wave function on the metal side with a decaying tail extending into the semiconductor side, implying these perturbed $p_z$-derived DB bands could be responsible for the conventionally hypothesized MIGS (as indicated by states F) since they share the same character in real-space distribution. In sharp contrast to the perturbed DB bands, the states of the unperturbed $p_x/p_y$-derived DB bands still retain their original DB character, decaying exponentially towards both sides, and eventually evolving into DBSS (as indicated by states A, C, and E). By examining the evolution of DB states across the three stages of Ag-Ge contact formation, it becomes clear that only the DB states involved in the localized bonding process across the interface during the first monolayer metal adsorption give rise to so-called MIGS in the MS contact upon deposition of a thick-layer metal (as indicated by states B, D, and F). In other words, MIGS originate specifically from the localized interface bonding between semiconductor DBs and metal orbitals at the interface. In the following, we will further demonstrate that such perturbed DB bands are the primary source of the conventionally postulated MIGS.

**B. Validate perturbed DB bands responsible for conventional MIGS**

The excellent agreement for thick-layer metal contact between our first-principles calculated SBHs and extensive experimental data, as summarized in Fig. 1, provides a robust foundation for validating our microscopic picture of interface states. Crucially, the calculated pinning factors ($S = 0$ for Ge and $S = 0.16$ for Si) are in quantitative agreement with their experimental counterparts ($S \approx 0.02$ for Ge and $S \approx 0.16$ for Si) across a wide range of metal work functions. This agreement with experimental measurements in such macroscopic properties confirms that our DFT-calculations correctly capture the total charge transfer and the interface dipole, which encompasses all interface states responsible for FLP, as illustrated in Fig. 2. In other words, there is no room for the existence of other types of gap states beyond those we have identified (i.e., the perturbed DB bands) to contribute to FLP at MS contacts. Prior works have also established that DFT reliably reproduces interface state behavior, including MIGS [10,31]. We can therefore safely conclude that the perturbed DB bands emerging from our calculations are the only source responsible for the MIGS, which was conventionally postulated to arise from the penetration of free-electron-like metal wave function into the semiconducting gap [38]. Furthermore, the close agreement in the pinning factor *S* between monolayer adsorption and thick-layer metal contact indicates that the interface dipole is predominantly determined by highly localized chemical bonding within the first atomic layer.

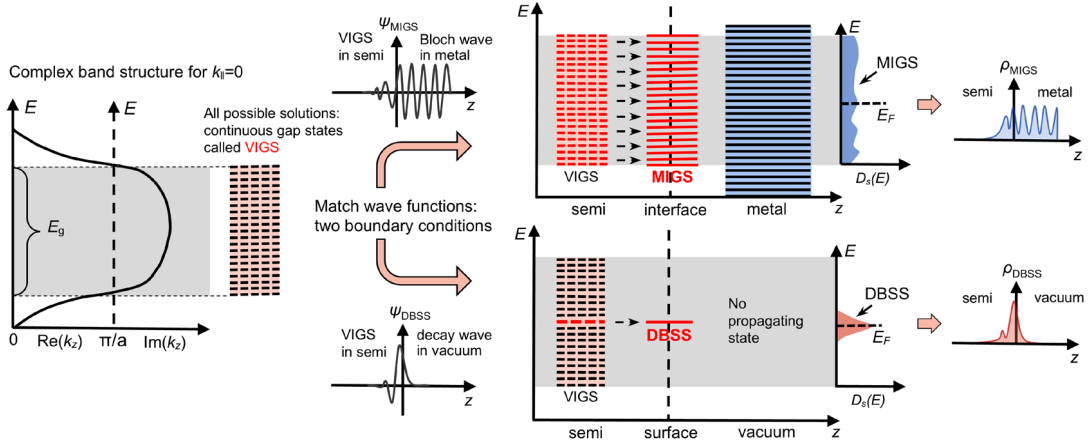

**FIG. 3.** Schematic illustration of the origin of DBSS and MIGS in a nearly-free-electron framework. Left panel: complex band structure $E(k_z)$ at a fixed in-plane wavevector $k_\parallel$, showing evanescent solutions (virtual-induced gap states, VIGS) with complex wavevector $k_z = \kappa + iq$. Middle panel: boundary condition matching at a semiconductor interface with metal (yielding broad MIGS) and with vacuum (yielding discrete DBSS).

To elucidate the physical origins of MIGS and DBSS, we employ a nearly-free-electron picture proposed by Heine, Tersoff, and Tung [1,6,7,19], which is naturally expressed in a plane-wave representation. As schematically illustrated in Fig. 3, for a fixed in-plane $k_\parallel$, the loss of periodicity along the z direction leads complex solution $k_z = \kappa + iq$ with energy levels inside the band gap, yielding evanescent states (often called virtual-induced gap states or VIGS) that decay exponentially into the semiconductor in the form $e^{-qz}$ [19,39,40]. While these evanescent states are valid solutions within the semiconductor side, they must also satisfy boundary conditions across the junction [19]. If the other side of the junction is metal, these evanescent solutions must match the propagating metal Bloch waves, producing extended MIGS with a broad energy distribution near $E_F$, consistent with the $p_z$-derived MIGS shown in Fig. 2(c). If the other side of the junction is a vacuum, the absence of propagating states on the vacuum side restricts the evanescent solution to one discrete DBSS per fixed $k_\parallel$ point for a given atomic orbital, resulting in a sharp peak of density of states (DOS) [1,19], as shown in Fig. 2(a). We note that, even at MS contacts, not all evanescent states find matching propagating metal Bloch waves. When boundary conditions forbid such matching, as observed for the $p_{xy}$-derived DB band in Ag-Ge contacts (Fig. 2(c)), DBSS emerges at the interface. While this nearly-free-electron model provides a universal foundation for how evanescent states evolve into MIGS or DBSS depending on the boundary conditions, a tight-binding model provides an equivalent but more intuitive framework for describing these localized interface states and directly linking them to the atomic-scale bonding structure, a connection that will be discussed later.

### C. The origin of MIGS and DBSS via symmetry analysis

By analyzing both the partial charge distributions and the energy state projections, we can distinguish MIGS from conventional DBSS to quantify their relative contributions to FLP in MS contacts (see Appendix A for the detailed calculation method). Specifically, Fig. 4(b) shows that the interface states of the Rh-Ge interface are dominated by MIGS, whereas the Ag-Ge interface exhibits both MIGS and DBSS, as shown in Fig. 4(c). The DBSS observed at the Ag-Ge interface closely resemble the original DB states of the bare Ge surface, as shown in Fig. 4(a), indicating that DBSS are mainly inherited from the original DB states. Fig. 4(d) and 4(e) show a high density of DBSS inside the Ge band gap for metal-Ge contacts with *s*- and *ds*-metals (such as Ag, Au and Mg), whereas for *d*-metals (such as Pt, Ir and Rh), the interface gap states are almost exclusively MIGS with few or even no DBSS. We attribute these variations primarily to differences in symmetry between the atomic orbitals of the metal and the dangling-bond orbitals of the semiconductor. According to group theory, coupling occurs only if metal and semiconductor states have the same irreducible representation, since for the coupling matrix element $V_{sm} = \langle \phi_s | V | \phi_m \rangle$ to be nonzero, the integrand must transform as the totally symmetric representation $A_1$ [41]. Because the symmetry of the unreconstructed Ge (001) $c(1\times1)$ bare surface is described by the $C_{2v}$ point group, the *s* and $p_z$ orbitals belong to the $A_1$ irreducible representation, while the $p_x$ and $p_y$ orbitals belong to the $B_1$ and $B_2$ representations, respectively. For *s*-metals and *ds*-metals, the states derived from *s* orbitals have $A_1$ symmetry and thus can effectively couple with the *s*- and $p_z$-type semiconductor DBSS (which are also $A_1$), but cannot couple with the $p_x$- and $p_y$-type DBSS (which are $B_1$ and $B_2$), leaving the latter largely unperturbed. Consequently, both MIGS and DBSS coexist in these MS contacts. In contrast, for *p*- and *d*-metals, the three-fold degenerate *p*-orbitals or five-fold degenerate *d*-orbitals decompose under the $C_{2v}$ point group into components having $A_1$, $B_1$, and $B_2$ symmetries, and thus can couple with all types of the semiconductor DBSS, giving rise to almost exclusively MIGS with negligible DBSS character, as shown in Fig. 4(d) and 4(e).

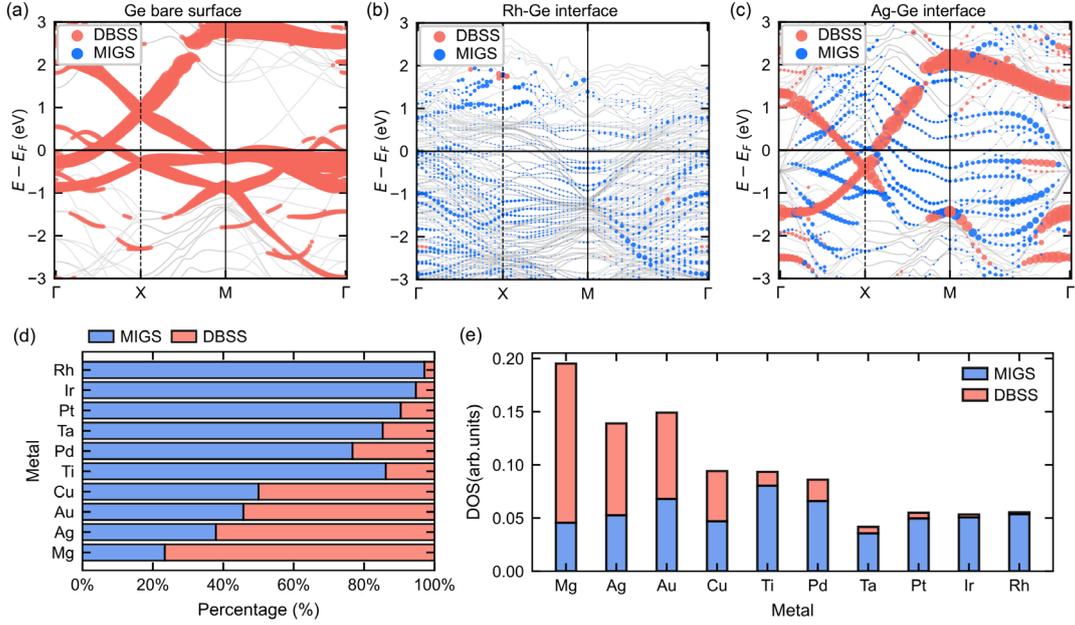

**FIG. 4.** First-principles calculated band structures for (a) the bare Ge surface, (b) thick metal Rh-Ge contact and (c) thick metal Ag-Ge contact. Blue and red circles represent the projections onto Ge surface atoms for MIGS and DBSS, respectively. (d) Relative contributions of MIGS and DBSS to the total interface states density and (e) the projected density of states (PDOS) onto the Ge surface atoms in different metal-Ge contacts, integrated over the entire band gap range.

Although the total number of gap states and the relative proportion of DBSS and MIGS vary among different metal contacts, it does not necessarily mean that an MS contact with fewer gap states (such as *p*- and *d*-metals in Fig. 4(e)) should exhibit weaker FLP with a larger pinning factor *S*. For instance, Fig. 5 shows that for single monolayer metals such as Pt, Rh, and Ir adsorbed on the Ge (001) surface, most gap states are shifted away from the band gap and down to within the valence band by strong metal-semiconductor bonding, leaving only a few within the band gap. In contrast,

Ag and Au monolayers show many perturbed DB states inside the band gap due to relatively weak bonding. However, these two sets of metals share the same FLP strength for both metal-Si and metal-Ge contacts. This is because interface states below the band gap still contribute to the interface dipole and, therefore, to the FLP strength. Upon increasing metal thickness, $s$- and $ds$-metal contacts preserve DBSS within the gap, whereas $p$- and $d$-metal contacts leave only a few in-gap bonding states that evolve into MIGS (Fig. 4(e)). Within the conventional gap states model, charge transfer is typically approximated as the filling of semiconductor gap states by metal electrons within the energy range between the Fermi level and CNL and is expressed as $eD_s(\Phi_{B,n} - \phi_{CNL})$. However, this approximation implicitly assumes that states below the valence band maximum (VBM) remain unchanged. When strong bonding drives many interface states deep into the valence band, these bonding states also accommodate electrons and thus contribute to the interface dipole. A formal derivation showing that this contribution is not negligible is provided in Appendix B.

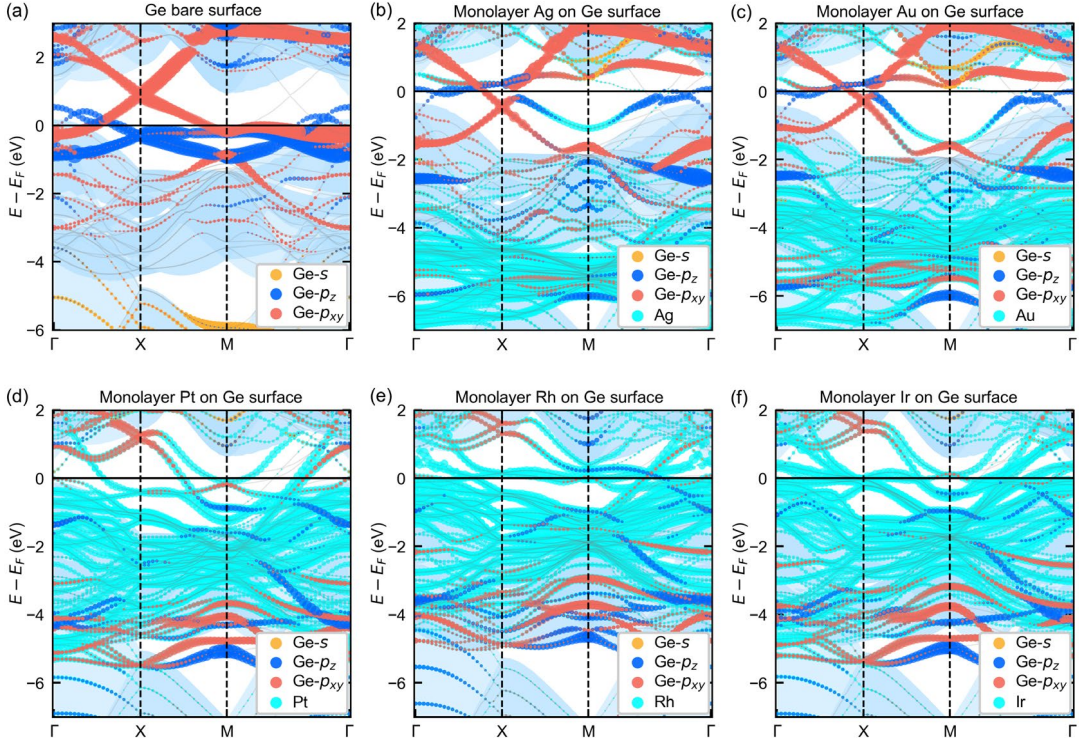

**FIG. 5.** First-principles calculated band structures of the bare Ge (001) surface and Ge (001) surfaces covered by monolayer metal atoms, over an energy range including both band gap and the valence band: (a) bare Ge (001) surface; (b) monolayer Ag; (c) monolayer Au; (d) monolayer Pt; (e) monolayer Rh; (f) monolayer Ir. The light-blue shadow represents the bulk Ge band structure

projected onto the (001) surface Brillouin zone, the gray lines denote the interface band structure and colored dots with varying sizes represent the contributions from Ge and metals.

### D. Charge transfer and interface dipole induced by localized interface bonding

After identifying DBSS, MIGS, and interface states in the valence band that originate from bonding between semiconductor DBs and metal orbitals at the MS interface, we now focus on the interface bonding process itself to clarify how it generates the interface dipole and thus the FLP effect. We begin by examining the charge transfer at the MS interface. Figure 6(a) compares the charge transfer, defined as the charge density difference $\Delta\rho(\mathbf{r}) = \rho_{\mathrm{Ag/Ge}}(\mathbf{r}) - \rho_{\mathrm{Ge}}(\mathbf{r}) - \rho_{\mathrm{Ag}}(\mathbf{r})$, at the metal-Ge interface for a thick Ag layer and for a single monolayer Ag on the metal side. The two cases exhibit a compelling similar spatial distribution, indicating that charge transfer is primarily determined by the coupling between the first metal monolayer and the semiconductor surface atoms. The bottom panel of Fig. 6(a) shows that the interface dipole potential $\Delta V$ obtained by solving Poisson's equation for $\Delta\rho(\mathbf{r})$ is responsible for FLP as described in Eq. (3). Therefore, the charge transfer induced solely by a single metal monolayer is sufficient to generate a large $\Delta V$, comparable to that produced by a thick-layer metal at the MS interface.

Figure 6(b) extends this analysis to other metals, showing similar parameters $k = \frac{d\Delta V}{d\phi_M}$ for both thick-layer and single monolayer metals contacts. These findings reinforce the concept that the interface dipole responsible for FLP arises primarily from a highly localized bonding process that occurs between semiconductor surface dangling bonds and metal surface atoms, accounting for roughly 90% of the total $\Delta V$ for all MS contacts. In the following section, we further demonstrate, from a tight-binding perspective, that localized interface bonding alone can indeed lead to similar large *k* and small *S*, in agreement with the DFT results. Similar conclusions have been reported by Tung, who showed that the potential drop across the MS interface is mainly confined to the surface metal atoms and the surface anion atoms of the compound semiconductor. We also show that, during the subsequent interface bonding process, some perturbed DB states evolve into MIGS, whereas others remain essentially unchanged. This suggests that the density of available DBSS on the bare semiconductor surface that participates in metal-semiconductor bonding is the decisive factor for FLP, whereas MIGS are not a determining factor themselves but instead arise as a secondary byproduct of this bonding process.

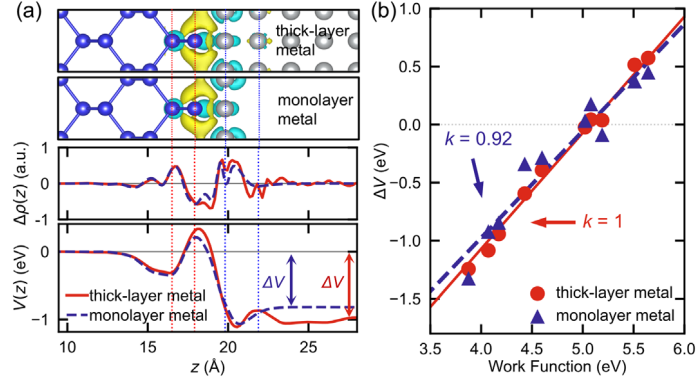

**FIG. 6.** (a) Charge density difference and interface dipole potential for thick-layer metal Ag-Ge and monolayer Ag-Ge interfaces. The top panel presents spatial distribution maps of the charge density difference $\Delta\rho(\mathbf{r})$, with yellow and blue isosurfaces indicating regions of charge accumulation and depletion, respectively. Blue and silver spheres represent Ge and Ag atoms, respectively. The middle panel shows the corresponding plane-averaged charge density difference $\Delta\bar{\rho}(z)$. The bottom panel displays the plane-averaged electrostatic potential $\bar{V}(z)$, obtained by solving Poisson's equation. The interface dipole potential $\Delta V$ is defined as the potential difference between two regions far from the interface, where the values converge to their respective bulk limits. (b) Relationship between the interface dipole potential $\Delta V$ and metal work function for both thick metal-Ge and monolayer metal-Ge interfaces.

**E. Tight-binding model for the formation of MIGS and DBSS originating from interface bonding**

To further verify correlation between DBSS and MIGS observed in the DFT calculations, we construct a toy model based on a one-dimensional atomic chain to mimic the coupling between DB states at the bare semiconductor surface and the atomic orbitals of the contacting metal in the context of the tight-binding approximation. Figures 7(a) and 7(b) schematically illustrate this toy model and its relevant parameters for two cases: contact with a single monolayer of metal and with multiple metal layers, respectively. For simplicity, both the semiconductor and the metal are represented by the same one-dimensional atomic chain with uniform lattice spacing. Since surface DB states within the band gap are strongly localized and closely resemble atomic orbitals (see Fig. 2), we use a single orbital with energy level $E_{DB}$ for each atom to represent the DB states at the semiconductor surface. In this simple model, we further set metal atomic orbital energy to $E_M = E_{DB} + 1$ and the coupling strength between nearest-neighbor metal atoms to $V_{mm} = 1$. The coupling strength $V_{sm}$ between the semiconductor DB state and the metal atomic orbital is then varied to investigate the resulting bonding states (Figs. 7(c)-(h)).

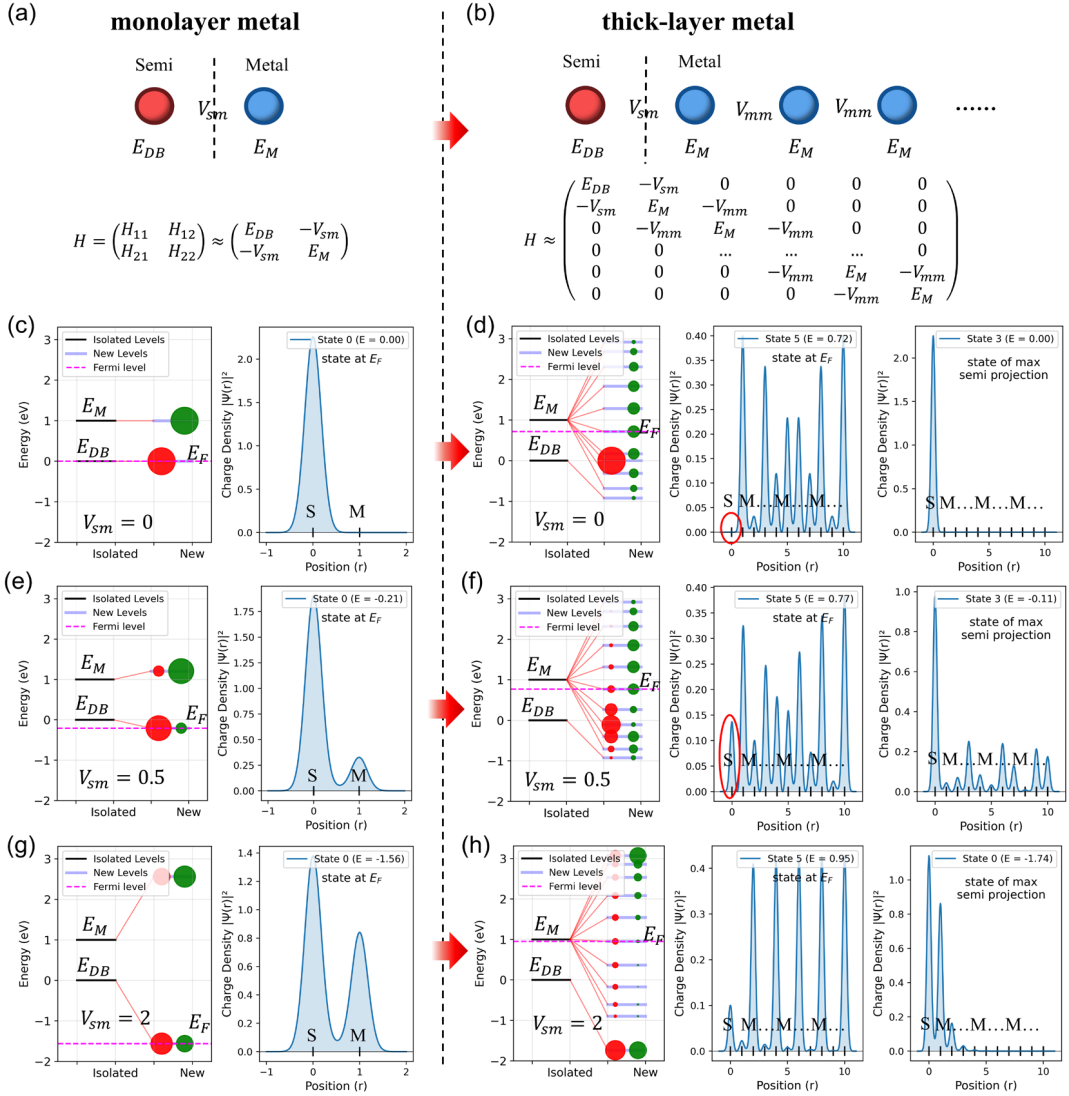

**FIG. 7.** Schematic illustration of the one-dimensional atomic chain tight-binding model and its simplified Hamiltonian for (a) a model with a single metal atom coupled to a semiconductor surface atom representing a DB and (b) a model extended to N (N > 1) metal atoms, forming a metal chain in contact with the semiconductor. Here, $E_{DB}$, $E_M$, $V_{sm}$, $V_{mm}$ denote the DB energy level, metal atomic energy level, coupling between the semiconductor and metal atom, and coupling between adjacent metal atoms, respectively. Subfigures (c)-(h) illustrate the shifts of the energy levels, as well as the orbital projections onto the semiconductor atom and the first metal atom in each new state (indicated by the size of the red and green circles, respectively). The schematic real-space charge density (represented using a Gaussian basis) is also shown for different cases, displaying only the new state nearest to the Fermi level or with the maximum projection on the semiconductor atom: (c) $N_m = 1$, $V_{sm} = 0$; (d) $N_m = 10$, $V_{sm} = 0$; (e) $N_m = 1$, $V_{sm} = 0.5$; (f) $N_m = 10$, $V_{sm} = 0.5$; (g) $N_m = 1$, $V_{sm} = 2$; (h) $N_m = 10$, $V_{sm} = 2$.

We first examine the scenario with no coupling, $V_{sm} = 0$, between the semiconductor DB state and the metal atomic orbital. In this limit, increasing the metal thickness from a single monolayer (N = 1) to 10 monolayers (N = 10) does not lead to the emergence of MIGS (Fig. 7(d)), because the bonding and antibonding states of the semiconductor DB state and the metal atomic orbital reduce simply to the original DB and metal orbitals, respectively (Fig. 7(c)). In this scenario, the charge density of the metal state vanishes exactly on the semiconductor side (see the middle panel in Fig. 7(d)), so there is no charge transfer across the interface and, consequently, no interface dipole arising from MIGS. The interface dipole thus stems exclusively from other metal electrons occupying the original half-filled DBs (right panel in Fig. 7(d)). When the coupling between semiconductor DB states and metal atomic orbitals is of moderate strength $V_{sm} = \frac{1}{2}V_{mm}$, new interface bonding states with bonding and antibonding character appear even for N = 1 (Fig. 7(e)). Upon increasing the metal thickness to N = 10, these new interface bonding states exhibit two distinct types of behavior: one type remains close to the original DB energy and behaves as interface resonance states (right panel in Fig. 7(f)), while the other type shows features of conventional MIGS, manifesting as bulk-like Bloch states on the metal side while decaying into the semiconductor side (e.g., states near the Fermi level shown in the middle panel of Fig. 7(f)). For very strong coupling $V_{sm} = 2V_{mm}$, highly localized bonding and antibonding states emerge with their energy levels significantly shifted downward and upward, respectively (Figs. 7(g) and 7(h)). These strongly bonded states move outside the metal band region, which explains the behavior observed for *d*-metal contacts in Fig. 5.

**F. Strong FLP explained by interface bonding within the bond-dipole model**

Building on the DFT results revealing the dominant contribution of monolayer-metal contacts to the interface dipole (Fig. 6), we now continue to use tight-binding approach to further elucidate how interface bonding gives rise to strong FLP and a large interface dipole. To capture this mechanism clearly, we start from the limiting case of N = 1 and employ Harrison's bond-orbital model [28], a particular implementation of the tight-binding approximation that uses hybridized bond orbitals (such as $sp^3$) as the basis set. In this framework, the Schrödinger equation for the system can be simplified as:

$$H\psi = \begin{pmatrix} \varepsilon_1 & -V_2 \\ -V_2 & \varepsilon_2 \end{pmatrix} \psi = E\psi \tag{7}$$

where $E_1$ and $E_2$ are the hybrid energies of $sp^3$ orbitals, and $V_2$ is the overlap integral, often referred to as covalent energy [28,42]. The evolution of eigenstates and the corresponding energy levels is summarized schematically in the left panel of Fig. 8(a). Increasing $V_2$ leads to a larger downward shift of bonding state (as also seen in Fig. 5). With one electron per orbital, the two electrons occupy the bonding state, resulting in a charge density difference:

$$\Delta\rho(\mathbf{r}) = \alpha_p |\phi_1(\mathbf{r})|^2 - \alpha_p |\phi_2(\mathbf{r})|^2 \tag{8}$$

where the polarity $\alpha_p = V_3/\sqrt{V_3^2 + V_2^2}$ and the polar energy $V_3 = (E_2 - E_1)/2$. Thus, $\alpha_p$ has a clear physical meaning: it quantifies the fractional electron transfer between the two sites, as illustrated in the left panel of Fig. 8(a).

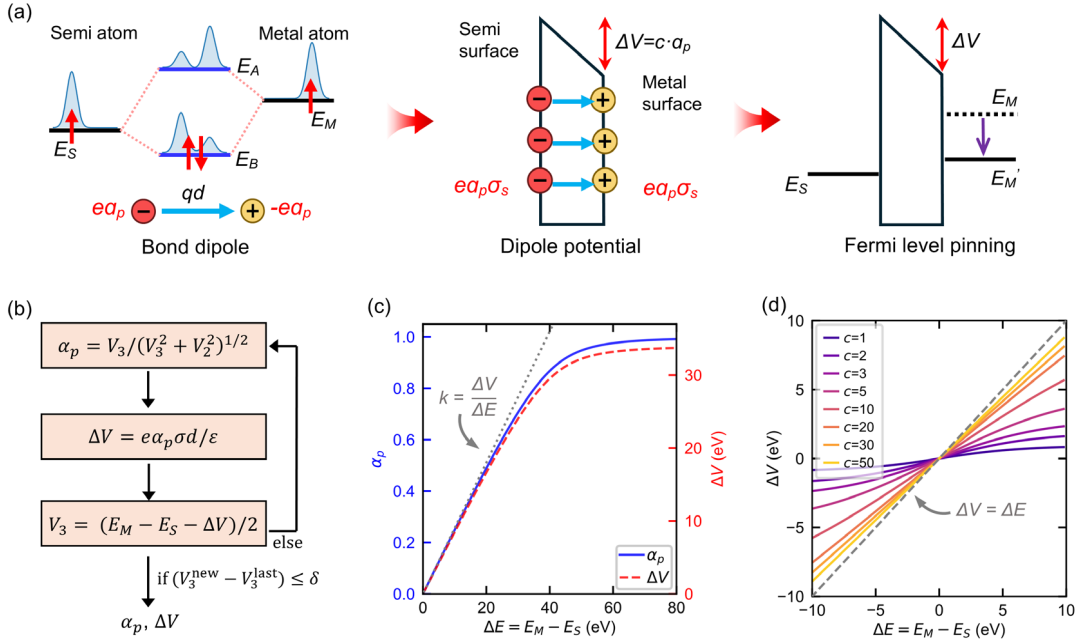

**FIG. 8.** (a) Schematic illustration of bond dipole formation at a metal-semiconductor interface. The formation of metal-semiconductor bonds gives rise to an interface dipole potential $\Delta V$, which drives the alignment of the two energy levels. $E_S$ and $E_M$ represent the energy levels of the semiconductor and the metal, respectively. The electron charge density associated with each energy state is also shown schematically. (b) Flow chart of the self-consistent calculation. (c) Calculated polarity $\alpha_p$ and interface dipole potential $\Delta V$ as a function of energy level difference $\Delta E = E_M - E_S$. The gray dotted line represents the linear approximation of $\Delta V$ near $\Delta E = 0$, indicating the initial slope $k = \Delta V / \Delta E$. Here, the structure of Ge (001) surface is considered: we take $\sigma_s = 4/a^2$ [28], and for simplicity use $d = 3$ Å [25,26], $\varepsilon_r = 2$ [10,43,44] and $V_2 = 3$ eV [28]. (d) Variation of the dipole potential $\Delta V$ for different scale factors $c$ in the typical energy level difference range $|\Delta E| < 10$ eV.

When two materials form an interface, the energy level difference between them drives charge redistribution and generates an interface dipole. Based on the bond dipole model shown in Fig. 8(a), electrons accumulate on the side with lower energy level, while the opposite side becomes positively charged, producing a potential drop:

$$\Delta V = \frac{Q \sigma_s d}{\varepsilon} = \frac{e \alpha_p \sigma_s d}{\varepsilon_0 \varepsilon_r} \equiv c \alpha_p \quad (9)$$

where $\sigma_s$ is a surface density of interface charge, $d$ is an effective dipole length describing the separation between the centers of positive and negative charge across the interface [25], and $\varepsilon_r$ is an effective interfacial dielectric constant [10,43,44]. The induced potential further modifies $V_3$ (see the right panel of Fig. 8(a)), which in turn affects $\alpha_p$, so that a self-consistent solution for $\Delta V$ is required, as illustrated in Fig. 8(b). Using representative parameters for the Ge (001) surface, where each (1×1) surface unit cell contains two surface atoms and each atom contributes two dangling-bond orbitals, so that $\sigma_s = 4/a^2$ ($a$ is the lattice constant) [28]. We find that as the energy level difference $\Delta E = E_M - E_S$ increases, the polarity $\alpha_p$ rises rapidly and saturates near 1, leading to a dipole potential $\Delta V$ that approaches the scale factor $c$ (Fig. 8(c)). The value of $c = \frac{e \sigma_s d}{\varepsilon_0 \varepsilon_r}$ critically determines how effectively the interface dipole compensates the initial energy misalignment. As shown in Fig. 8(d), a larger $c$ yields a larger $\Delta V$ over the entire range of $\Delta E$, and drives the slope $k = \frac{\partial(\Delta V)}{\partial(\Delta E)}\Big|_{\Delta E \to 0} \approx \frac{\Delta V}{\Delta E}$ towards 1. This corresponds to strong FLP ($k \to 1$ and $S \to 0$), where the semiconductor hybrid level acts as the CNL [14]. Thus, localized interface bonding can generate an interface potential drop that tends to compensate the initial energy level misalignment. In addition, the close agreement between monolayer and thick-layer contacts in Fig. 1 and Fig. 6(b) further indicates that adding more metal layers only weakly modifies the interface dipole, and that a single monolayer already captures its dominant contribution.

Finally, within this unified interface bonding framework, we can also understand why more ionic semiconductors generally exhibit weaker pinning than covalent ones, as demonstrated in our previous study [30]. In highly ionic semiconductors, surface states formed upon cleavage have orbital character and energy close to the bulk band edges, typically near the conduction band minimum (CBM) or VBM rather than deep within the gap as in covalent semiconductors. These orbitals are highly localized and couple only weakly with metal wave functions. Moreover, these surface states are typically either fully occupied or empty, and lie energetically far from the metal Fermi level, which further suppresses charge transfer. Since we have shown that the key factor controlling FLP is the number of DBs on the bare surface that can form new bonds with the metal, which is represented by the surface density parameter $\sigma_s$ in $c = \frac{e\sigma_s d}{\varepsilon_0 \varepsilon_r}$, the density of such available DB orbitals on ionic semiconductor surface is much lower. This leads to a smaller $c$ and hence a smaller $k$ (as shown in Fig. 8(d)), in agreement with our previous finding [30]. It also implies that FLP can be tuned by reducing the density of available surface DBs before metal deposition, for instance, by increasing surface ionicity through appropriate adsorbates or by passivating the more spatially extended DBs arising from hybridized orbitals on covalent semiconductor surfaces.

## III. CONCLUSIONS

In summary, by combining first-principles calculations with a tight-binding analysis, we demonstrate that all newly generated states, including MIGS, DBSS, and bonding states embedded in the valence band, can be understood within a unified bond dipole framework for interface dipole formation, as they arise from the same underlying mechanism. In this picture, unperturbed dangling bonds residing in the band gap behave as metal-semiconductor bonding states with very weak coupling to the metal ($\alpha_p \to 0$); in other words, they retain a more ionic character and therefore do not shift downward into the valence band. In contrast, MIGS originate from the broadening of DB states in the gap that are perturbed by the metal. We further show that the very initial stage of interface formation, with chemical bonding highly localized to the outermost interface atomic layer, is already sufficient to induce strong FLP. By formulating a bond dipole theory for MS interfaces based on Harrison's bond-orbital model, we clarify the microscopic origin of interface dipole and provide a unified description of FLP across different material systems. Overall, our results offer a coherent microscopic framework for understanding the electronic properties of MS interfaces and provide practical guidance for interface engineering.

**ACKNOWLEDGMENTS**

The work was supported by the National Natural Science Foundation of China (NSFC) under Grant No. 12525402, and the CAS Project for Young Scientists in Basic Research under Grant No. YSBR-026.

**Appendix A: Distinguish DBSS from MIGS and quantify their contributions**

The classification of MIGS and DBSS is based on a combination of fat-band (band structure projected onto selected atoms) calculations and partial charge density analysis. The band structure is projected onto three distinct regions: the first surface layer of semiconductor atoms (region 1), the central bulk region of metal atoms (region 2), and the central bulk region of semiconductor atoms (region 3). According to the different characteristics of MIGS and DBSS, states located inside band gap that have negligible projection weight (defined as the norm of the wave function projected onto the selected region) on region 2 but finite weight on region 1 are classified as DBSS, while states with finite projection weight on both regions 1 and 2 are classified as MIGS. States with substantial weight primarily on region 1 or region 3 but lying below the band gap are attributed either to semiconductor bulk states or to surface states resonant with bulk states. To ensure the robustness of this classification, we cross-check each state using its corresponding charge density distribution (partial charge density analysis). To quantitatively estimate the gap states density and the relative contributions of MIGS and DBSS, we average the local density of states (LDOS) over the entire band gap range to mitigate fluctuations.

**Appendix B: Clarify the role of deep valence-band bonding states**

The conventional gap states model posits that only interface states lying between the CNL and Fermi level contribute to charge transfer, leading to a simplified expression [20,26]:

$$Q = D_s(\Phi_{B,n} - \phi_{\text{CNL}}). \tag{B1}$$

The main limitation of this model is its exclusive focus on dipole contributions originating from states inside the band gap. Our analysis suggests that many of the bonding states below the VBM can be traced back to strongly perturbed DB bands that have shifted further downward, suggesting a common origin with the perturbed DBSS remaining in the gap. Therefore, to accurately describe the interface dipole, the gap states theory should be generalized to include all occupied states below $E_F$, including those below the VBM in the charge transfer calculation. In the simplest one-dimensional case, the charge transfer $Q$ (defined as the additional charge density localized on the semiconductor surface after interface formation) can be expressed as:

$$Q = q^{MS} - q^S = \int_{E_{\min}}^{E_F^{MS}} D_S^{MS}(E)\, dE - \int_{E_{\min}}^{E_F^S} D_S^S(E)\, dE \tag{B2}$$

where $q^{MS}$ and $q^S$ are the charge densities on the semiconductor side near the surface in the MS contact and in the bare semiconductor surface, respectively; $E_F^{MS}$, $E_F^S$, are the corresponding Fermi levels; and $D_S^{MS}(E)$ and $D_S^S(E)$ are the semiconductor's surface density of states for the MS contact and the isolated surface, respectively [20]. By splitting the integration range, Eq. (B2) can be rewritten as:

$$Q = \left[\int_{E_{\min}}^{E_F^S} D_S^{MS}(E)\, dE + \int_{E_F^S}^{E_F^{MS}} D_S^{MS}(E)\, dE\right] - \int_{E_{\min}}^{E_F^S} D_S^S(E)\, dE$$

$$= \int_{E_{\min}}^{E_F^S} [D_S^{MS}(E) - D_S^S(E)]\, dE + \int_{E_F^S}^{E_F^{MS}} D_S^{MS}(E)\, dE \tag{B3}$$

If we further assume: (1) $D_S^{MS}(E) = D_S^S(E)$ over the entire energy range of $E_{\min} \leq E \leq E_F^S$, and (2) $E_F^S = \phi_{\text{CNL}}$ with an energy-independent $D_S^{MS}(E)$, Eq. (B3) reduces to $D_S^{MS}(E_F^{MS} - \phi_{\text{CNL}})$, which is equivalent to Eq. (B1). However, as discussed in Fig. 2 and Fig. 5, interface bonding inevitably modifies the energy level distribution of the semiconductor surface, so the condition $D_S^{MS}(E) = D_S^S(E)$ holds only in the absence of metal-semiconductor coupling ($V_{sm} = 0$). Under realistic conditions with significant coupling, this assumption breaks down, and the first integral in Eq. (B3) generally cannot be neglected. Therefore, Eq. (B1) describes only an idealized situation with no interface interaction.

**References**

[1] V. Heine, Theory of Surface States, Phys. Rev. **138**, A1689 (1965).

[2] C. R. Crowell *et al.*, Attenuation Length Measurements of Hot Electrons in Metal Films, Phys. Rev. **127**, 2006 (1962).

[3] W. Mönch, On metal-semiconductor surface barriers, Surf Sci. **21**, 443 (1970).

[4] N. F. Mott, The theory of crystal rectifiers, Proc. R. Soc. London, Ser. A **171**, 27 (1939).

[5] W. Schottky, Zur Halbleitertheorie der Sperrschicht- und Spitzengleichrichter, Z. Phys. **113**, 367 (1939).

[6] J. Tersoff, Schottky Barrier Heights and the Continuum of Gap States, Phys. Rev. Lett. **52**, 465 (1984).

[7] R. T. Tung, The physics and chemistry of the Schottky barrier height, Appl. Phys. Rev. **1**, 011304 (2014).

[8] T. Nishimura, Understanding and Controlling Band Alignment at the Metal/Germanium Interface for Future Electric Devices, Electronics **11**, 2419 (2022).

[9] T. Nishimura, K. Kita, and A. Toriumi, Evidence for strong Fermi-level pinning due to metal-induced gap states at metal-germanium interface, Appl. Phys. Lett. **91**, 123123 (2007).

[10] S. G. Louie, J. R. Chelikowsky, and M. L. Cohen, Ionicity and the theory of Schottky barriers, Phys. Rev. B **15**, 2154 (1977).

[11] W. E. Spicer *et al.*, New and unified model for Schottky barrier and III–V insulator interface states formation, J. Vac. Sci. Technol. **16**, 1422 (1979).

[12] R. T. Tung, Chemical bonding and Fermi level pinning at M-S interfaces, Phys. Rev. Lett. **84**, 6078 (2000).

[13] O. F. Sankey, R. E. Allen, and J. D. Dow, Theory of Schottky barrier formation for transition metals on Si, Ge, diamond, and Six Ge1−x alloys, J. Vac. Sci. Technol. B **2**, 491 (1984).

[14] W. A. Harrison and J. Tersoff, Tight-binding theory of heterojunction band lineups and interface dipoles, Journal of Vacuum Science & Technology B: Microelectronics Processing and

Phenomena **4**, 1068 (1986).

[15] A. M. Cowley and S. M. Sze, Surface States and Barrier Height of Metal-Semiconductor Systems, J. Appl. Phys. **36**, 3212 (1965).

[16] J. Bardeen, Surface States and Rectification at a Metal Semi-Conductor Contact, Phys. Rev. **71**, 717 (1947).

[17] J. Ihm, S. G. Louie, and M. L. Cohen, Diamond-Metal Interfaces and the Theory of Schottky Barriers, Phys. Rev. Lett. **40**, 1208 (1978).

[18] W. Mönch, Role of virtual gap states and defects in metal-semiconductor contacts, Phys. Rev. Lett. **58**, 1260 (1987).

[19] H. Lüth, *Solid surfaces, interfaces and thin films* (Springer, Berlin, 2001).

[20] J. Ihm, S. G. Louie, and M. L. Cohen, Electronic structure of Ge and diamond Schottky barriers, Phys. Rev. B **18**, 4172 (1978).

[21] T. Iffländer *et al.*, Local Density of States at Metal-Semiconductor Interfaces: An Atomic Scale Study, Phys. Rev. Lett. **114**, 146804 (2015).

[22] R. T. Tung, Schottky-Barrier Formation at Single-Crystal Metal-Semiconductor Interfaces, Phys. Rev. Lett. **52**, 461 (1984).

[23] W. A. Harrison, Theory of band line-ups, Journal of Vacuum Science & Technology B: Microelectronics Processing and Phenomena **3**, 1231 (1985).

[24] W. R. L. Lambrecht and B. Segall, Theory of Semiconductor Heterojunction Valence-Band Offsets: From Supercell Band-Structure Calculations toward a Simple Model, Phys. Rev. Lett. **61**, 1764 (1988).

[25] W. R. L. Lambrecht and B. Segall, Interface-bond-polarity model for semiconductor heterojunction band offsets, Phys. Rev. B **41**, 2832 (1990).

[26] R. T. Tung, Formation of an electric dipole at metal-semiconductor interfaces, Phys. Rev. B **64**, 205310 (2001).

[27] R. T. Tung and L. Kronik, Fermi level pinning for zinc-blende semiconductors explained with interface bonds, Phys. Rev. B **103**, 085301 (2021).

[28] W. A. Harrison, *Elementary electronic structure* (World Scientific, Singapore, 2004).

[29] A. Dimoulas *et al.*, Fermi-level pinning and charge neutrality level in germanium, Appl. Phys. Lett. **89**, 252110 (2006).

[30] Z. Xiang, J.-W. Luo, and S.-S. Li, Self-passivation Causes the Different Fermi Level Pinning between Metal-Si and Metal-Ge Contacts, arXiv preprint arXiv:2411.14220 (2024).

[31] S. G. Louie and M. L. Cohen, Self-Consistent Pseudopotential Calculation for a Metal-Semiconductor Interface, Phys. Rev. Lett. **35**, 866 (1975).

[32] Ç. Güven and N. Ucar, Schottky barrier height dependence on the metal work function for p-type Si Schottky diodes., Zeitschrift für Naturforschung A **59a**, 795 (2004).

[33] X. Luo *et al.*, Understanding of Fermi level pinning at metal/germanium interface based on semiconductor structure, Appl. Phys. Express **13**, 031003 (2020).

[34] R. R. Lieten *et al.*, Ohmic contact formation on n-type Ge, Appl. Phys. Lett. **92**, 022106 (2008).

[35] T. Nishimura, T. Yajima, and A. Toriumi, Reexamination of Fermi level pinning for controlling Schottky barrier height at metal/Ge interface, Appl. Phys. Express **9**, 081201 (2016).

[36] S. M. Sze, Y. Li, and K. K. Ng, *Physics of semiconductor devices* (John Wiley & Sons, Hoboken, 2007).

[37] H. B. Michaelson, The work function of the elements and its periodicity, J. Appl. Phys. **48**, 4729 (1977).

[38] M. Kobayashi *et al.*, Fermi level depinning in metal/Ge Schottky junction for metal source/drain Ge metal-oxide-semiconductor field-effect-transistor application, J. Appl. Phys. **105**, 023702 (2009).

[39] V. Heine, Some theory about surface states, Surf. Sci. **2**, 1 (1964).

[40] P. Mavropoulos, N. Papanikolaou, and P. H. Dederichs, Complex Band Structure and Tunneling through Ferromagnet/Insulator/Ferromagnet Junctions, Phys. Rev. Lett. **85**, 1088 (2000).

[41] P. Y. Yu and M. Cardona, *Fundamentals of Semiconductors* (Springer, Berlin, Heidelberg, 2010).

[42] D. G. Pettifor, *Bonding and Structure of Molecules and Solids* (Oxford University Press, New York, 1995).

[43] J. P. Walter and M. L. Cohen, Wave-Vector-Dependent Dielectric Function for Si, Ge, GaAs, and ZnSe, Phys. Rev. B **2**, 1821 (1970).

[44] R. Resta, Thomas-Fermi dielectric screening in semiconductors, Phys. Rev. B **16**, 2717 (1977).